\documentclass[runningheads]{llncs}

\usepackage[utf8]{inputenc}
\usepackage{graphicx}
\usepackage[dvipsnames,table,xcdraw]{xcolor}

\usepackage{verbatim}
\usepackage{xcolor,colortbl}
\usepackage{pgfplotstable,filecontents}
\pgfplotsset{compat=1.9}
\usepackage{enumitem}
\usepackage{url}
\usepackage{ulem}
\usepackage{breakurl}
\usepackage[breaklinks]{hyperref}
\hypersetup{colorlinks=true,urlcolor=blue}
\usepackage[numbers,sort&compress,square]{natbib} 
\usepackage{wrapfig}
\usepackage{longtable,tabu}

\usepackage{multirow}
\usepackage{multicol}
\usepackage{booktabs}
\usepackage{threeparttable}
\usepackage{textcomp,amsmath,longtable}
\usepackage{latexsym}
\usepackage{caption}
\usepackage{subcaption}
\usepackage{hyperref}
\usepackage[colorinlistoftodos,prependcaption, textsize=small]{todonotes}

\usepackage{rotating}

\usepackage{lipsum}

\usepackage{amsmath}
\usepackage{floatrow}

\newcommand{\smalltilde}{\raise.17ex\hbox{$\scriptstyle\mathtt{\sim}$}} 
\newcommand{\s}{\texttt{sim}}

\usepackage{fancyhdr}
\usepackage{float}
\floatstyle{plaintop}
\restylefloat{table}
\usepackage[tableposition=top]{caption}

\title{A Mathematical Framework for Evaluation of SOAR Tools with Limited Survey Data
\thanks{\scriptsize{This manuscript has been co-authored by UT-Battelle, LLC, under contract DE-AC05-00OR22725 with the US Department of Energy (DOE). The US government retains and the publisher, by accepting the article for publication, acknowledges that the US government retains a nonexclusive, paid-up, irrevocable, worldwide license to publish or reproduce the published form of this manuscript, or allow others to do so, for US government purposes. DOE will provide public access to these results of federally sponsored research in accordance with the DOE Public Access Plan (\url{http://energy.gov/downloads/doe-public-access-plan}).}}
} 
\author{Savannah Norem\orcidID{0000-0002-6289-077X} \and
Ashley E Rice \orcidID{0000-0001-7560-0017} \and 
Samantha Erwin\orcidID{0000-0002-5162-4193} \and
Robert A Bridges \orcidID{0000-0001-7962-6329} \and 
Sean Oesch \orcidID{0000-0002-6909-1022} \and
Brian Weber \orcidID{0000-0002-3261-5152}}
\institute{Oak Ridge National Laboratory, Oak Ridge, TN 37830, USA \\ \email{noremsa@ornl.gov}}

\begin{document}
\authorrunning{S. Norem et al.}
\titlerunning{A Mathematical Framework for Evaluating SOAR Tools}
\maketitle
\vspace{-7mm}
\begin{abstract}
Security operation centers (SOCs) all over the world are tasked with reacting to cybersecurity alerts ranging in severity. Security Orchestration, Automation, and Response (SOAR) tools streamline cybersecurity alert responses by SOC operators. 
SOAR tool adoption is expensive both in effort and finances. Hence, it is crucial to limit adoption to those most worthwhile; yet no research evaluating or comparing SOAR tools exists. The goal of this work is to evaluate several SOAR tools using specific criteria pertaining to their usability. SOC operators were asked to first complete a survey about what SOAR tool aspects are most important. Operators were then assigned a set of SOAR tools for which they viewed demonstration and overview videos, and then operators completed a second survey wherein they were tasked with evaluating each of the tools on the aspects from the first survey. In addition, operators provided an overall rating to each of their assigned tools, and provided a ranking of their tools in order of preference. Due to time constraints on SOC operators for thorough testing, we provide a systematic method of downselecting a large pool of SOAR tools to a select few that merit next-step hands-on evaluation by SOC operators. Furthermore, the analyses conducted in this survey help to inform future development of SOAR tools to ensure that the appropriate functions are available for use in a SOC. 
\keywords{SOAR Tools \and User Study \and Cybersecurity.}

\end{abstract}
\vspace{-7mm}
\section{Introduction and Background}
\vspace{-4mm}
\label{sec:intro} 
The term \textit{security operations center} (SOC) refers to the subteam of an organization's IT department tasked with maintaining the network's cyber health---the confidentiality, integrity, and availability of the enterprise's data and systems.
SOCs are now equipped with a widespread collection of data from a vast array of sensors feeding logs, alerts, and raw data to a security information and event management system (SIEM). 
SIEMs and back-end infrastructure generally provide SOCs customizable, real-time dashboards and rapid data query. 
However, the process of identifying incidents, responding appropriately, and documenting findings remains to be done by the operators. The SIEM, along with the many other SOC tools, can only provide limited awareness, leading to extended threat detection and response times and reduced ability to prevent or quickly mitigate an attack \citep{Network, islam_babar_nepal_2019}. 

Further, SOC efficiency suffers due to time spent documenting findings, collecting evidence across an array of sources, and completing other administrative-type procedures \citep{BREWER20198}. Security orchestration, automation, and response (SOAR) tools---the term coined by Gartner in 2017 \citep{BREWER20198}---are the newest generation of software tools designed to enable SOC operators to more efficiently and uniformly detect and address cybersecurity threats. 
While orchestration and automation are often marketed together, they do have distinct functions and can be distinguished \citep{islam_babar_nepal_2019} as follows: \textit{orchestration} specifically refers to the integration of separate tools with different functions into a single platform to streamline and accelerate the investigation of a threat; \textit{automation} refers to reducing the manual effort required by SOC operators during investigation and threat response phase \citep{Microsoft}. 
SOAR tools aim to help by automating parts of an investigation and helping SOC operators prioritize alerts to investigate. 
A SOAR tool has the following defining capabilities: 
(1) ingests a wide variety of SOC data, 
(2) assists in prioritizing, organizing, and displaying the data to the users, 
(3) allows customizable workflows or ``playbooks'' to standardize SOC procedures, 
(4) provides automation to expedite the SOC's procedures, 
(5) integrates with a ticketing system, and 
(6) facilitates collaboration of operators, potentially in disparate geographic locations or networks. 
The capabilities of a SOAR tool are observed with the integration of a SOAR mechanism to secure energy microgrids---wherein the tool collects and contextualizes security data from multiple sources in the microgrid, performs an overall sweep of the system to identify present vulnerabilities, and initialize or engage a response to detected threats \citep{microgrid}.

SOAR tools are a transformational and centerpiece tool for a SOC promising measurable gains in effectiveness and efficiency; 
as such, they require substantial investments. 
In addition to monetary costs, configuration depends on standardizing and codifying workflows, and incurring technological and organizational debt. 
Hence, efforts to find the right SOAR tool for a SOC are warranted, yet because SOAR tools are so new, there is little research assessing their usability or the degree to which they improve SOC operators' ability to respond to threats. 
This work will provide an in-progress report on the first-ever user study to assess and compare user preferences of SOAR tools with the aim of downselecting to a smaller number of tools to be extensively tested. 
Our goal is to provide scientific assistance to an organization to winnow the wide variety of SOAR tools to the best one for the organization's needs. 

A first step for evaluating SOAR tools, and the focus of this work, is down-sampling the variety of tools to a subset worthy of the costly hands-on evaluation without discarding a potentially desired tool. 
We describe both our experimental design methodology and anonymized results---as a prerequisite, SOAR vendors agreed to providing licenses and support for this study in exchange for anonymity of their results and participation---of actual SOC operators evaluating 11 market-leading SOAR tools based on vendor-supplied technological overviews. Previous work has used game theory to aid in decisions of mid-size enterprises on cybersecurity tools \cite{decision_support}, or conducted user studies questioning users why they do (not) employ specific cybersecurity measures \cite{use_non_use}. Yet none of this work addresses how users perceive current market leading tools that are available to them.



In terms of understanding usability of SOAR tools, this work aids in understanding what defining capabilities are most desired by SOC operators and how this affects their preferences. 
Our unique pipeline provides a method for narrowing down to only the tools that are most preferred by SOC operators. 
This pipeline's many unique and modular components could be applicable to future user studies, in particular: 
statistical simulations to quantify confidence of results under varying number of participants;
a novel algorithm for optimizing participant assignments; 
a framework for optimizing combinations and variants of multi-criteria recommender system (for predicting missing ratings) to the data at hand that outperforms the previous versions; 
an evaluation of supervised regressors for predicting overall ratings from aspect ratings is provided (and verifies that this is a needed step as it greatly out-performs predictions from the recommender system); 
many methods for ranking the tools based on the data are provided side-by-side including a novel application of PageRank to convert ratings data to ranking data; 
statistical analyses that investigate correlations present. 
To assist the community in reusing any of these methods, code will be released \href{https://github.com/noremsa/SurveyAnalysisFramework}{here} once open-source copyright is gained. 

In summation, this work will present a novel pipeline of ingesting sparse data and using a mathematically sound method to fully populate the matrix of data. Novel methods are then applied to create multiple views of data along with the statistical relevance of demographic factors. This paper will fill a gap in knowledge where we start with a survey of what the users of these tools actually want, and will proceed to an in depth user study of what the tools deliver. We first present our methods, beginning with study design and progressing through our algorithms and machine learning techniques for handling missing data and predicting overall ratings. We then detail our use of PageRank for rank aggregation, along with statistical methods to analyze demographic impact on user responses. 
\section{Methodology}
\vspace{-4mm}
In total, 11 SOAR tools are included in this study.
Before the study, operators were asked to provide an ordered ranking of the defining capabilities (listed in the SOAR definition above) in order of importance. 
Participants watched two vendor-prepared videos, one that gives of their tool and one that provides a demonstration. We provided general guidelines regarding the videos, wherein the overview must include information on the technical approach of the tool (architecture, deployment, algorithms, etc.) as well as the novelty of the tool. We requested that the demonstration video provide an introduction to the platform with examples of users viewing events, using playbooks, collaborating, along with capabilities to automatically populate tickets and orchestrate multiple incidents. Participants then provided per-capability Likert scale ratings (1 to 5), an overall tool rating, and ranked all tools viewed. 
\vspace{-2mm}
\subsection{Data Collection}
\vspace{-2mm}
\subsubsection{Survey Design.}
\label{sec:survey-design}
Prior to the survey in which the operators evaluated the tools in this study, we collected information from the operators about which defining capabilities of SOAR tools are most important to them. 
Four SOCs participated in this preliminary conversation and informed the questions we asked in our Internal Review Board \footnote{The IRB works to ensure the rights of participants in any human subject study are fully protected} approved full survey.
The capabilities in question included: the ability to rank/sort alerts so that high priority alerts are emphasized, ability to automate common tasks, easy of playbook creation and modification, ability to provide a unified experience across geographic locations, ability to ingest disparate data sources, and ability to pre-populate alerts and tickets with additional context. 
These same aspects were the criteria by which the operators evaluated the SOAR tools. For each question, answers followed the 1-5 Likert scale, with an additional option for ``Can't Tell." 

Before the set of SOAR tool videos and surveys were administered, a set of demographic questions were posed to each participant including how long they had been in their role, what their role was, the level of familiarity they had with each specific SOAR tool, and finally, asking them to rank by importance the defining capabilities of a SOAR tool. 

After viewing all SOAR tool videos and completing the per-tool surveys, users were asked to rank the tools they had reviewed. The survey was provided online via a secured website to protect the sensitive nature of both SOAR tool privacy videos as well as information provided by SOC analysts. 
The full survey can be found in Table \ref{tab:Survey}. 

\subsubsection{How many participants to recruit?} 
\begin{table}[!b]
 \centering
\resizebox{\textwidth}{!}{%
\setlength\extrarowheight{4pt}
\begin{tabular}{p{1.3cm}lcccccc}
\toprule
\textbf{\#reviews/pair:} & & \textbf{10} & \textbf{15} & \textbf{25} & \textbf{30} & \textbf{35} & \textbf{40} \\ 
\midrule
\multirow{4}*{\rotatebox{90}{Different}} 
 & $z$ test & (0.88, 0.96) & (0.89, 0.99) & (0.95, 1.0) & (0.97, 1.0) & (0.98, 1.0) & (0.98, 1.0) \\
 & $\chi^2$ test & (0.04, 0.23) & (0.06, 0.37) & (0.17, 0.71) & (0.24, 0.81) & (0.29, 0.88) & (0.34, 0.91) \\
 & Bin. test & (0.36, 0.68) & (0.42, 0.8) & (0.57, 0.93) & (0.63, 0.97) & (0.66, 0.97) & (0.71, 0.99) \\
\midrule
\multirow{4}*{\rotatebox{90}{Same}}
& $t$-test & 0.04 & 0.05 & 0.05 & 0.03 & 0.05 & 0.06 \\
& $z$-test & (0.73, 0.82) & (0.79, 0.86) & (0.81, 0.89) & (0.83, 0.89) & (0.86, 0.91) & (0.86, 0.91) \\
& $\chi^2$-test & (0.0, 0.02) & (0.0, 0.01) & (0.01, 0.03) & (0.01, 0.01) & (0.01, 0.02) & (0.02, 0.02) \\
& Bin.-test & (0.1, 0.18) & (0.13, 0.18) & (0.19, 0.19) & (0.2, 0.21) & (0.2, 0.17) & (0.23, 0.21) \\
\bottomrule
\end{tabular}} 
 \captionsetup{font = scriptsize}
\caption{For each simulated number of participants, the table reports the percent of the simulations for which the hypothesis test confirmed the two tool's data were different. For example, a result of 0.67 implies that test provided at least 95\% confidence that the tool 2 rating was different from the tool 1 rating in 67\% of the simulations. For the $z$-, $\chi^2$-, and Binomial tests, Likert values $\{1, \dots, 5\}$ are converted to binary $\{0,1\}$ using thresholds 2.5, then 3.5. Both results are given in the form of (\%\{thresh = 2.5\}, \%\{thresh = 3.5\}). 
Since these tests were testing the null hypothesis that the ratings were sampled from the same distribution, higher (lower) percentages in the top (bottom) half when the distributions were different imply high (low) performance by the test, respectively.}
\label{tab:sim_table}
\end{table}
In general, unless the differences between the groups are rather extreme, small sample sizes do not yield enough statistical power. We begin this study by choosing an appropriate sample size to ensure validity of the results. We design and run a simple (two-tool, single-rating) simulation to quantify the sensitivity of statistical significance to number of participants. 
Although our actual user study will include several tools and ratings, this simple exercise will be sufficient to provide quantifiable reasoning about the number of participants to target. 

For the simulation, we consider two scenarios in which a pair of tools are rated on the  Likert scale: (1) a pair of tools are rated differently (tool 2 preferred to tool 1) and (2) a pair of tools are rated the same (no preference for tool 1 over tool 2, or vice versa).
We sample $m = 5, 10, \dots, 40$ participants' overall ratings of the two tools---one overall rating sampled per user per tool--from two distributions over sample space $\{1, 2, \dots, 5\}$. We then compare results of different hypothesis tests that assess whether the ratings are sampled from the same distribution.

First, we examine the number ($m$) of pairwise tool reviews needed to have confidence that tool 2 is preferred over tool 1 given distribution means of $\mu = 3.65$ and $\mu = 2.93$, and variances of $\sigma^2 = 1.17$ and $\sigma^2 = 0.923$ for tool 2 and tool 1, respectively. 
Second, we examine the number of reviews needed to tell with confidence that tool 2 is preferred equally to tool 1 by using the same distribution ($\mu = 3.65, \sigma^2 = 1.17$) for both. 
In both scenarios, we run the simulation 100 times and for each compute a two-sided $t$ test. We use Welch's $t$ test assuming variances are not equal for the first scenario. 

Next we convert the Likert data to binary data using a threshold of 2.5 such that ($1,2 \mapsto 0; 3,4,5 \mapsto 1$), and with a threshold of 3.5 where ($1,2,3 \mapsto 0; 4,5 \mapsto 1$). 
We compute the $z$-score on difference of means, a $\chi^2$ test with Yates' correction on the difference in proportions of 0 and 1s, and two binomial tests (first hypothesis is tool 2 is sampled from tool 1's distribution, second is vice-versa) for each. 
The statistical tests in this section were designed based on Loveland \cite{Loveland} and applications of Sauro \& Lewis \cite{Sauro_Lewis_2012}. 
 
Our results (Table \ref{tab:sim_table}) confirm that more reviews provide higher, or at least negligibly worse, percentages of correct conclusions in all cases. 
Our results show that the $z$ test and Binomial test under both thresholds are poor at identifying when tools are the same, whereas the $\chi^2$ test with threshold of 2.5 is poor at telling when they are different. 
However, the $t$ test has good performance in both scenarios, as does the $\chi^2$ test with threshold of 3.5, so we use these results for our target number. 
These estimates are based only on the specific distributions used in the two scenarios (which may not match real-world data we obtain), but are reasonable assumptions to provide a quantitative approach to reasoning how many operators are needed. 

We presented these statistical results to the sponsor organization allowing them to quantifiably reason about the balance between their operator's time and statistical power of the results desired. Note that there are 11 total tools, and thus [11 choose 2] = 55 unique pairs of tools, and secondly that each participant will be asked to rate eight of 11 tools to respect their time, yielding [8 choose 2] = 28 pairwise reviews. Thus, 1,650 desired total desired pairwise reviews divided by 28 pairwise reviews per participant yields a target of 59 participants. After being presented with this information along with Table \ref{tab:sim_table} our sponsoring organization decided that nineteen participants was sufficient. 
\subsubsection{How to assign reviews to each participant?} 
Since we only will require 8 of 11 tools to be reviewed by each participant, the problem of which 8 tools to assign to each participant arises because the assignment of tools to reviewers affects how many pair reviews we obtain. 
We seek to maximize the number of reviews for every pair of tools given a fixed number of participants. 
Further, the algorithm must accommodate more/less participants than is desired to accommodate optimistic/realistic participation. 

Note that independent of the assignment, for $n$ total tools, $m$ participants, and $l$ reviews per participant there will be an average of $\mu := m \times [l$ choose $2]/[n$ choose $2]$ ratings of each pair of tools. 
Our assignment algorithm seeks to find $m$ different $l$-tuples of tools so that
each pair of tools occurs in as close to $\mu$ of the tuples as possible (each pair is assigned to exactly $\mu$ participants). 
For our case ($m = 59, l = 8, n = 11)$ we seek $\mu = 59 \times 28/55 \sim 30$ reviews for each pair. 

For each $l$-tuple, we enumerate the $[l$ choose $2]$ pairs in the $l$-tuple, and for each pair, we track the number of times that tuple has already occurred. 
Initially all counts are 0, and we define the score for the $l-$tuple as the sum of these tallies. 
While the set of assignments is less than $m$, the algorithm sorts all $l-$tuples by score, and picks the next participant's assignment uniformly at random from those $l-$tuples with minimal score, then updates the tallies and scores of all $l$-tuples. 
Once an assignment is given (a list of $m$ total $l$-tuples), we define the assignment error as the average absolute difference of the number of pairwise assignments from $\mu$. 
Although this algorithm may not find an optimal assignment, it is usually close.
We set $m$ larger than the desired number of users, to optimistically accommodate over-recruitment, and run the algorithm $100$ times (10s), keeping only the best assignment. 
Since the recruitment in practice yielded only 19 participants, we note that because the algorithm is greedy, using the first 19 assignments is also close to optimal. 
In our case of $n = 11, m = 59, l = 8$, we have $\mu = 30.0\overline{36}$, and the resulting assignment produced 6 pairs with 29, 41 pairs with 30, and 8 pairs with 31 reviews for an average error of $\sim0.28$. 

\subsubsection{Sentiment Analysis}
After watching the videos for each tool, participants were asked to complete the free response question, ``Is there anything else about this tool that you would like to share?" 
The problem for analyzing this feedback is it is text, i.e., not numerical. 
In order to convert text feedback into a numerical rating, our algorithm takes the average of two sentiment analyzers, VADER\footnote{\url{https://www.nltk.org/_modules/nltk/sentiment/vader.html}} \cite{hutto2014vader}, which provides a polarity score based on heuristics leveraging sentiWordnet \cite{baccianella2010sentiwordnet}, and roBERTa\footnote{\url{https://huggingface.co/cardiffnlp/twitter-roberta-base-sentiment}} \cite{barbieri2020tweeteval}, a sentiment analyzer that uses BERT \cite{devlin2019bert} and is trained on sentiment-labeled tweets to produce a polarity score.
The average of these scores provides a polarity (in $[-1,1]$), which we map to $[1,5]$ Likert scale.
Finally, we included a check for similarity of the VADER and roBERTa scores, manually inspecting if they differed more than .5, which never occurred herein. 
Overall, the free responses are parsed into strings and converted to a Likert value using sentiment analysis, which can be analyzed similar to the other numeric feedback. 
\vspace{-2mm}
\subsection{Predicting Missing Ratings}
\vspace{-2mm}
Since operators recruited for this survey were assigned a subset of the tools to review, this means that we have missing data since every operator did not complete the survey for each tool. 
This section provides a linear progression of previous research that informs our approach followed by our algorithm for filling in missing values to populate a complete dataset. 
Our contributions include providing a simple Bayesian technique for computing similarities dependent on unseen ratings and a method to optimally weight multiple predicted ratings, which, at least on our data, outperforms previous methods. 
Consider a vector, $\vec{r}(u, t, :)$, that comprises the ratings assigned by user $u$ to tool $t$; i.e., $\{\vec{r}(u,t,i)\}_{u,t,i}$ is a tensor, $\vec{r}$ with the third dimension representing the ratings.

\subsubsection{Relevant Recommender System Literature.} 
Breese et al. \cite{breese1998empirical} considers the single-criteria problem where $\vec{r}(u,t, :) = r(u,t)$ is a scalar, equiv. $r$ is a matrix. Unknown values are defined as follows, 
\begin{equation}
 \label{eq:breese}
 \vec{r}(u,t,:): = \frac{1}{\sum_{v\in \text{Users}} \s(u,v) } \sum_{v\in \text{Users}} \s(u,v) \vec{r}(v,t,:)
\end{equation}
where $\s(u,v)$ is a similarity measure of users $u$ and $v$ computed with a standard similarity measure (e.g, cosine, inverse Euclidean, Pearson correlation) applied to the vector of ratings from $u$ and $v$ on the set of items both users rated. 

Adomavicius \& Kwon (AK) \cite{adomavicius2007new} extend this framework to multi-criteria ratings where $\vec{r}(u,t)$ is a vector with $\vec{r}(u,t,n)$ user $u$'s overall rating of tool $t$ on the $n^{th}$ aspect. 
Setting a distance on rating vectors, $d_R(\vec{r}(u,t,:), \vec{r}(v,s,:))$, 
AK defines the distance between two users, $u, v$, as \\
$
 d_{U}(u,v) = \sum_{t\in T_{u,v}} d_R(\vec{r}(u,t), \vec{r}(v,t,:)) /|T_{u,v}|,
$
where $T_{u,v}:= \{$tools $t$ rated by both $u$ and $v$\}. The similarity of two users is simply an inverse function of their distance such that $\s(u,v):= 1/(1+d_{U}(u,v))$. 
Notably, the general framework is symmetric in users and items; hence, item similarity could just as easily produce predicted missing values. 
Finally, for each missing aspect rating ($j = 1, ..., n-1$), the output $\vec{r}(u,t,j)$ is predicted
according to Eg. \ref{eq:breese}, supervised learning techniques are suggested for learning $\vec{r}(u,t,n)$ from the aspect ratings $\{\vec{r}(u,t,j): j = 1, ..., n-1\}.$
 The progression of the research literature diverged from these ``instance-based'' methods to develop ``model-based'' methods, e.g., 
 \cite{hofmann1999latent, si2003flexible, sahoo2012research}, which seek Bayesian network models that include latent variables designed to encode clusters of similar users and of items.
Results of Sahoo et al. \cite{sahoo2012research} conclude that model-based methods excel (in precision-recall balance and in precision-rank balance) when $\vec{r}$ is sparse (common, e.g., in online marketplaces with a huge inventory of items), whereas, the instance-based method of AK excels
for dense $\vec{r}$, which is the case in this study. 

\subsubsection{Predicting Aspect Ratings.}
For each user, $u$, and for each tool, $t$, we have a vector of numeric responses $\vec{r}(u,t,:)$ of length eight, with the first seven entries as the aspect ratings--- the six questions on the SOAR tool's capabilities and the numerically-converted text comments field---and the last entry the overall rating by user $u$ for tool $t$.
Hence, $\vec{r}$ is a stack of eight $19\times11$ matrices. 
As each user was assigned a minimum of eight tools (of 11 total), on average we expect 3 tools $\times$ 8 ratings missing, leaving a whole vector $\vec{r}(u,t,:)$ empty. 
In addition to empty entries due to how tools were assigned, some entries were empty due to an operator leaving a question blank for a tool that they were assigned to evaluate. 
In all there were approximately one third missing values. 
Following the results of Sahoo et al. \cite{sahoo2012research}, we use the AK workflow with tailored modifications. 

From our data we define and compute three different similarities. Each produces predictions for all unknown values following Eq. \ref{eq:breese}, and we learn an optimal convex combination of the three as our prediction. 
The first two similarities are simply the user similarity $\s_U(u,v)$ and tool similarity $\s_T(s,t)$ from the ratings matrix.
To compute these, we first need a distance on the rating vectors. 
We use $\ell^p$ distance and test four distances, $p = 0, 1, 2, \infty$, where $p=0$ denotes counting the number of unequal entries in the two input vectors.
For our implementation, we define 
$\s_U(u,v) := \exp{(-\| \vec{r}(u,:,:) - \vec{r}(v,:,:)\|_p)}$, and 
$\s_I(s,t) := \exp{(-\| \vec{r}(:,s,:) - \vec{r}(:,t,:)\|_p)}$. 

When computing $\s_U$, the original work of AK ignored items that were not rated by both users. 
We tested both the naive rating distances against a Bayesian version. 
To compute the Bayesian distance between two vectors that may be missing values, we simply marginalize over a uniform distribution on the set of all possible missing values, (uniform on $\{1,...,5\}$).
Upon training we will have eight ratings distances to consider parameterized by $p \in \{0, 1, 2, \infty\}$ and each using the naive or Bayesian approach.

Recall from Section \ref{sec:survey-design} our survey asked each participant to rank the aspects of a SOAR tool in advance of seeing and rating any of the tools.
This is valuable information for understanding each user's preferences, and we take this into account by providing a second user similarity (third similarity in total) from this data, which in turn provides a prediction of unknown ratings. 
We simply define $\s_{U_{rank}}(u,v)= (1 + \texttt{kendalltau}(u,v) )/2$ where the function \texttt{kendalltau} computes the Kendall Tau correlation \cite{knight1966computer} of the two users' aspect rankings. 
As $\text{kendalltau}(u,v) \in [-1,1]$ this similarity achieves its minimum of -1 with opposite rankings input, and maximum of 1 with identical rankings input. 

Let $\vec{r}_U, \vec{r}_I,$ and $\vec{r}_{U_{rank}}$ denote the populated tensors with previously missing values now predictions from $\s_U, \s_T, \s_{U_{rank}}$, respectively. Define the unknown ratings as $\vec{r}: = (1-a-b) \vec{r}_U + b \vec{r}_I + a \vec{r}_{U_{rank}}$, where $a \in [0,1]$, and $b \in [a, 1]$ are weight parameters to be learned. 
To learn the parameters, we grid search over $a, b, p$ and across using naive vs. Bayesian ratings distances.
For each set of parameters, we compute the macro-averaged error over a 20-fold cross validation to find the most accurate combination. 
We use 20-fold cross validation so that each fold has only 5\% known but hidden values used for testing.

Our optimal parameters were found to use the Bayesian ratings distance computation with $p=1$, $a = 0.2$, and $b = 0.2$, which exhibited the (lowest) average error of 0.682. 
Since we used a grid search, the previous methods (non-Bayesian distances, using only a single similarity) are included in the results, and hence our advancements to provide greater accuracy than previous methods. 

\subsubsection{Predicting overall ratings.}
While overall ratings were included in the predictions above, we suspect (as AK \cite{adomavicius2007new} suggests) that overall ratings are more accurately predicted from the seven aspect ratings (for that tool by that user). 
To this end, we test a wide variety of supervised machine learning algorithms for regressing the overall ratings from the corresponding learning algorithms on a held out test set. 

After creating fully populated ratings for each aspect of each tool for each user, we first take the data that has a user-given overall ranking. We use this populated data to train and test ten machine learning regression algorithms. We then do a five-fold cross-validation on each model to determine which has the smallest error. We use the results from the model that produced the lowest average mean squared error (MSE) across the five folds. The model with the lowest MSE is then trained on all the available user given rankings. We then use the predicted aspect ratings to predict the overall rating.
\vspace{-2mm}
\subsection{PageRank}
\vspace{-2mm}
Using the predicted overall rating, we developed a directed graph. Each vertex of the directed graph represents a tool. The edges between each node are drawn with an arrow, where the arrow points to higher rated tools based on each user. For example, if a user rated Tool A: 5; Tool B: 3; Tool C: 1, there would be directed edges as follows:
$C \rightarrow B, \quad
C \rightarrow A, \quad
B \rightarrow A. $
In cases where a user rates tools with the same number, no edges are drawn between those tools that would indicate which one is higher rated. In the case of duplicate links, the edges become weighted. A single edge has a weight of 1, if the directed edge is duplicated then 1 is added to the weight for every duplicate. If the converse directed edges is added, then 1 is deducted from the weight of the edge. This method was particularly applicable because we considering filling in missing data in a pairwise fashion, and this algorithm compares the tools pairwise.

After developing a weighted directed graph, we use the PageRank algorithm to measure the importance of each node \cite{page1999pagerank}. To implement the PageRank algorithm we used the NetworkX PageRank link analysis toolbox \cite{hagberg2008exploring}. 
\vspace{-2mm}
\subsection{Statistical Analyses}
\vspace{-2mm}
When we interpret results, we need to consider them in the context of certain demographics if it is shown that these factors have an impact on how operators are rating these SOAR tools. Through these univariate and multivariate analyses, we sought to determine the strength of four relationships: 
\begin{enumerate}
\item Tool rating and operator experience (in years)
\item Tool rating and operator occupation (security operator or other)
\item Tool rating and operator familiarity with the tool
\item Tool rating and perceived video quality by the operator
\end{enumerate}
We employed three tools to conduct these analyses: linear regression, Kruskal-Wallis Test, and multivariate analysis of variance (MANOVA). Each of these methods are described briefly below.

\subsubsection{Linear Regression.}
For each tool, we regressed its overall rating onto years of experience the operator has. After a line is fit to the data, Wald's Chi-Squared Test is to determine whether years of experience has an impact on the rating of the tool. Here, the null hypothesis is that the slope of the line that we fit to the data is zero, indicating that there is no relationship at all between years of experience and tool rating. Based on a 5\% error rate, a sufficiently small $p$-value ($<$ 0.05) from Wald's Chi-Squared Test indicates that the slope of the best fit line is \textit{not} zero and there is likely a relationship between the variables. 
\subsubsection{Kruskal-Wallis.}
In our analysis of the impact of tool familiarity on tool rating, we sought to address the question ``Do users assign higher ratings to tools with which they have more familiarity?'' Familiarity had five categories that users could mark for the tools they reviewed: \textit{Currently use it often, Used it at least once, Used it often in the past, Heard of it, Never heard of it}. 

For this analysis, we use a non-parametric Kruskal-Wallis test to compare the median overall tool scores of each of these five groups. All tool scores from every group are put into a single vector and sorted in ascending order. The tools are then ranked by their position in order from 1...$n$, where $n$ is the number of observations. For each of the five group, a sum of the ranks from each of its observations are calculated. These sums, along with the sample size and number of groups, are used to calculate an $H$ statistic. The $H$ statistic then gets compared to a critical chi-squared value at a certain error rate. Should the $H$ statistic be greater than the critical value, the null hypothesis (``the medians of these five groups are the same'') is rejected in favor of the alternate hypothesis (``the medians of the five groups are not the same'').

A Kruskal-Wallis Test was also performed to answer the question ``Does perceived video quality impact tool ratings?'' After watching the overview and demonstration videos about the tools, users specified whether they felt the video was \textit{Great, Okay, Terrible}. 

\subsubsection{MANOVA.}
We use a MANOVA test to determine whether a user's occupation impacts how they rated tools, and whether certain occupations preferred specific tools. MANOVA is useful for testing the effects that one explanatory variable (occupation) has on two or more dependent variables (ratings of tools 1-11) and compares the means of multiple dependent variables across two or groups. 
\section{Results}
\vspace{-4mm}
\subsection{Results from Raw Data}
\vspace{-2mm}
Survey results were collected from four different SOCs, for a total of 19 SOC operators and 158 reviews (10-17 reviews per tool). The survey had two sections: evaluation of what users wanted most out of a SOAR tool (\textit{aspect ratings}), evaluation of how users scored each of their assigned SOAR tools on these aspects, and how they ranked them in order of preference (\textit{tool ratings and rankings).} 
\subsubsection{Tool results}
Each SOC operator watched a tool overview video and a tool demonstration video. Following completion of the video reviews, the operators completed our survey that asked questions about each  aspect of the tool, provided an overall tool ranking (1: most preferred), and then rated (scored) all of the tools (1: lowest score).

\begin{wrapfigure}[13]{r}{0.5\textwidth}
    \vspace{-10mm}
    \centering
    \includegraphics[scale = 0.45]{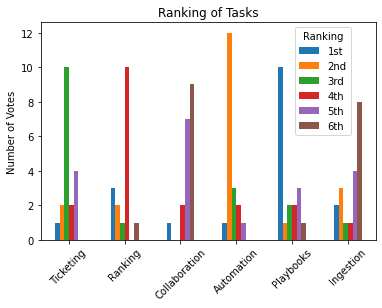}
    \captionsetup{font=scriptsize}
    \captionsetup{width=\textwidth}
    \vspace{-10pt}
    \caption{Each aspect's rank of importance from 1 (very important) to 6 (not important)} 
    \vspace{-10pt}
    \label{fig:taskranking}
\end{wrapfigure}
Before investigating individual tools, we asked each operator to rank six aspects, or tasks that a SOAR tool could perform, in order of descending importance. In Figure \ref{fig:taskranking}, we plot how many times each tool was ranked in each position. In this grouped bar chart, we see that playbooks are ranked first by ten operators, indicating that operators prioritize playbooks and workflows that are easy to manage. In second place, most operators voted that automating common tasks was important, followed by pre-populating tickets in third and ranking alerts in fourth.  Fifth and sixth places are less clear, but we can conclude the last two places are reserved for collaboration ability and ingestion of data from various sources.

\subsubsection{Aspect Ratings.}
Before investigating individual tools, we asked each operator to rank six aspects, or tasks that a SOAR tool could perform, in order of descending importance. In Figure \ref{fig:taskranking}, we plot how many times each tool was ranked in each position. In this grouped bar chart, 
we see that playbooks are ranked first by ten operators, indicating that operators prioritize playbooks and workflows that are easy to manage. In second place, most operators voted that automating common tasks was important, followed by pre-populating tickets in third and ranking alerts in fourth.  Fifth and sixth places are less clear, but we can conclude the last two places are reserved for collaboration ability and ingestion of data from various sources.

\begin{wrapfigure}[17]{l}{0.65\textwidth}
    \vspace{-25pt}
    \includegraphics[scale = 0.32]{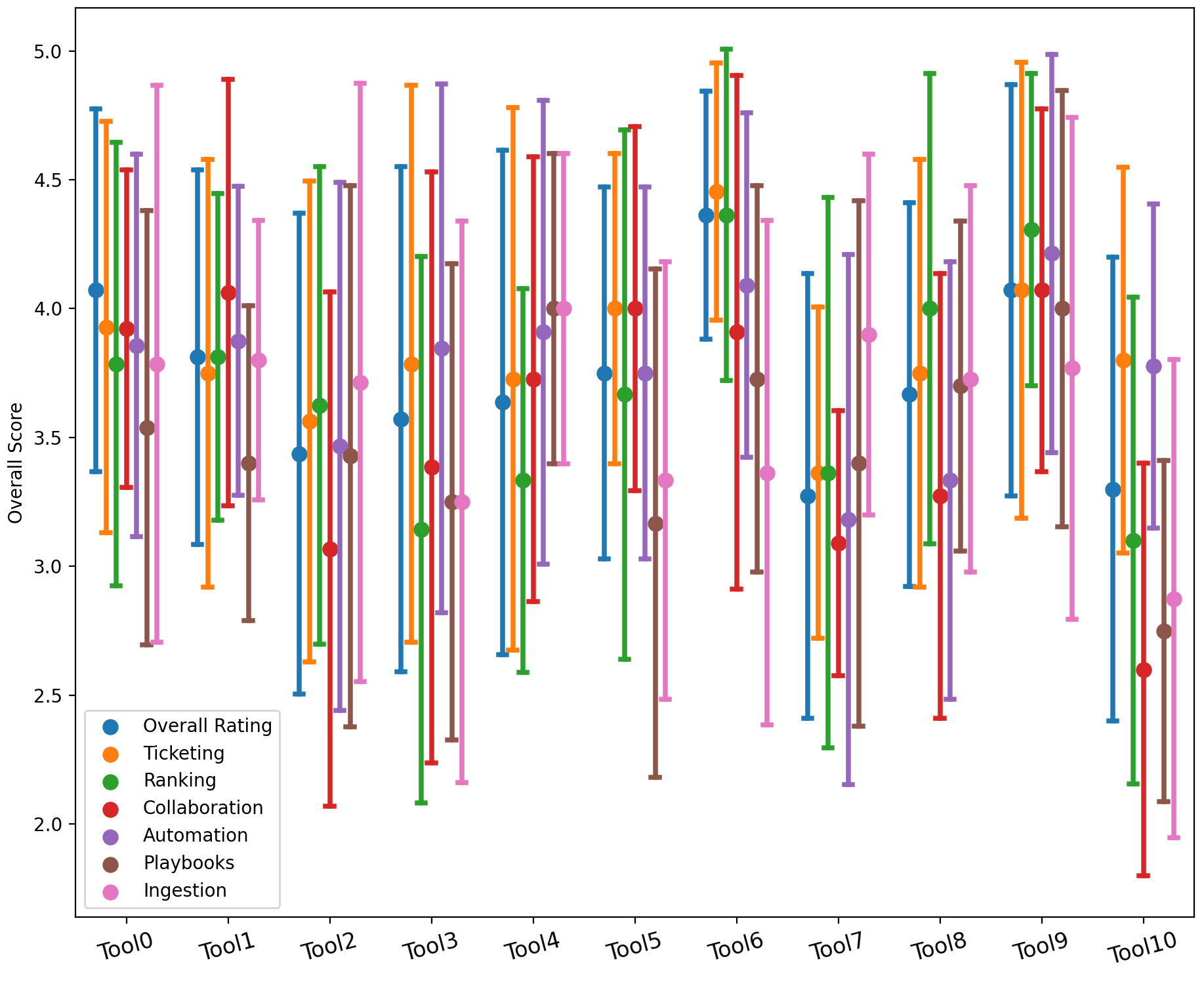}
    \captionsetup{font = scriptsize}
    \vspace{-10pt}
    \caption{Summary of all of the raw survey scores. Each tool is on the X-axis, and the score is on the Y-axis. The unique colors represent a specific aspect of the survey.} 
    \label{fig: RawScore}
\end{wrapfigure} 

In Figure \ref{fig: RawScore}, we note that Tool 6 is scored the highest on average with a mean rating of 4.36 out of 5. Contrarily, Tool 7 is rated the lowest on average with a mean rating of 3.27.  Operators ranked playbook management as the most important aspect (Figure \ref{fig:taskranking}) and the tools received an average score of 3.56 on this aspect. Tool 9 performed the best on their playbook aspect, and scored a 4.07, while Tool 10 performed the worst.  Operators rank automation as the second most important feature of a SOAR tool, and we found that on average our tools received a rating of 3.75. We again see Tool 9 perform the best, and Tool 7 perform the worst.

\begin{wrapfigure}[13]{r}{0.45\textwidth}
    \vspace{-13mm}
    \centering
    \includegraphics[scale = 0.3]{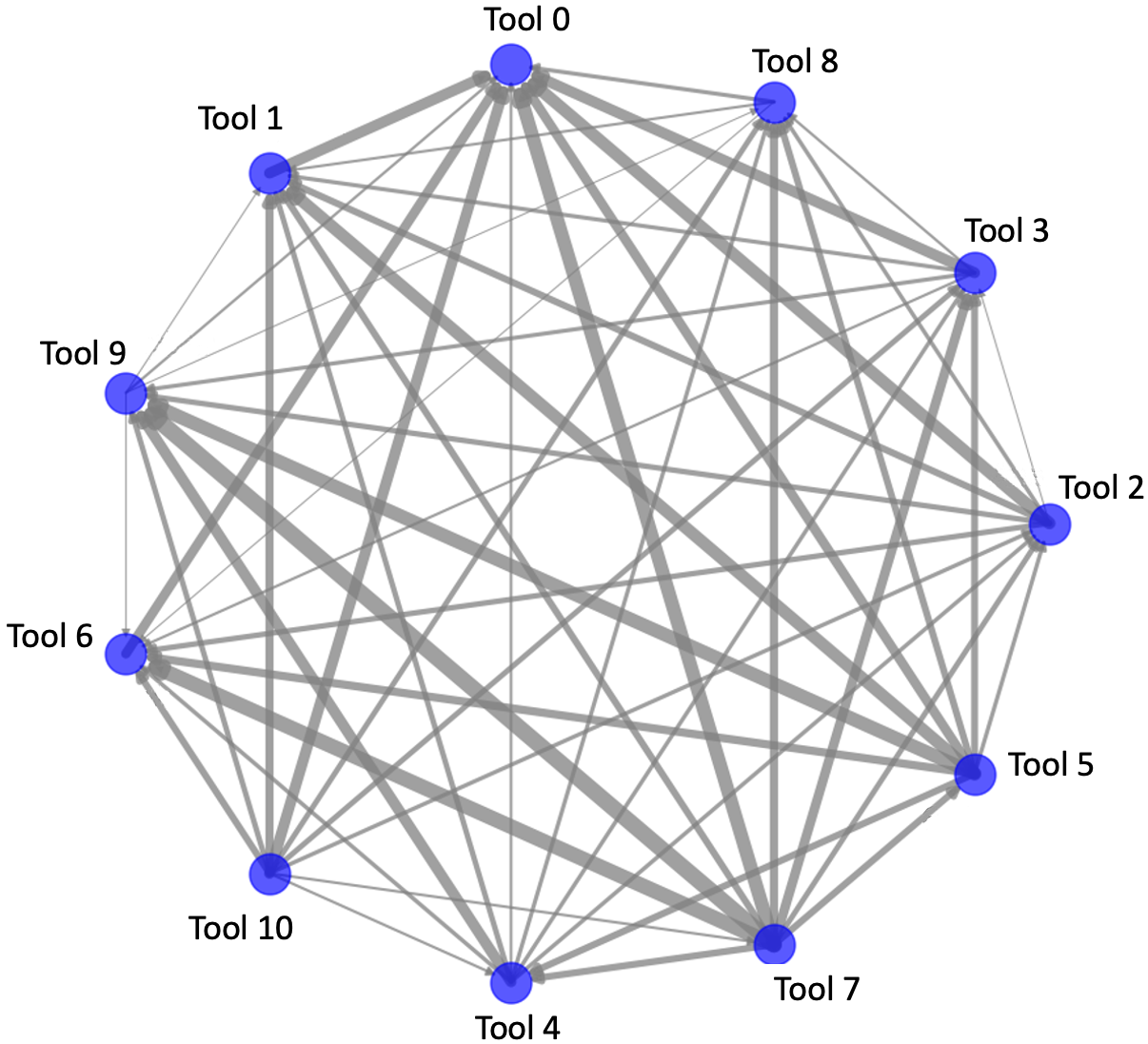}
    \captionsetup{font = scriptsize}
    \captionsetup{width=0.8\textwidth}
    \caption{PageRank graphical network from raw data.}
    \label{fig:RawPR}
\end{wrapfigure}
We performed a PageRank analysis on the tool rankings to identify the most preferred tool based on the user's rankings alone, given the seemingly inconsistent ratings and rankings of the tools. The directed PageRank graph (Figure \ref{fig:RawPR}) is based on user rankings, and we note that Tool 6 has the most inward weights, which indicates that most users ranked this tool as the best. Analogously, we see Tool 0 has the most outward weights, indicating that most users ranked this tool last.
\vspace{-2mm}
\subsection{Results from Populated Data}
\vspace{-2mm}
The following results are derived from the populated data, or the data for which we have filled in all the missing aspect ratings and used these to predict the missing overall ratings. With the populated data, we have a total of 197 complete tool reviews. 

\subsubsection{Predicted Responses}
\begin{figure}[H]
    \centering
    \includegraphics[scale = 0.45]{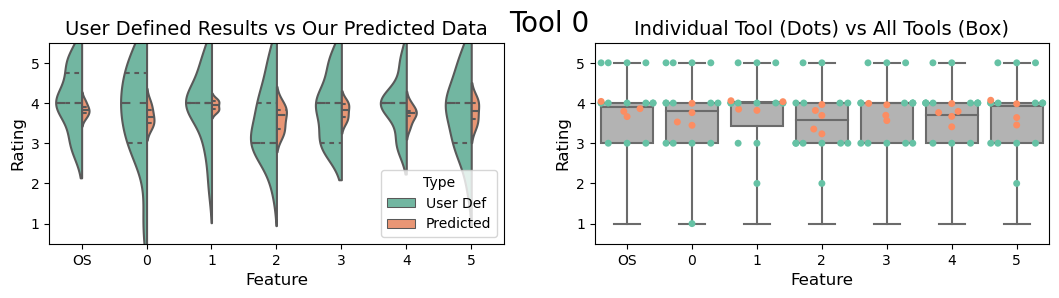}
    \captionsetup{font = scriptsize}
    \caption{Tool 0 score summary based on user defined results and predicted results. In the left panel, the distribution of the user defined results (green) and the predicted results (red). In the right panel, comparison of the results (green/red dots, respectively) with the average scores across all tools (grey box plot).}
    \label{fig:toolEx}
\end{figure}

 In Figure \ref{fig:toolEx} we highlight how our predicted responses compare the user defined responses in this single example of Tool 0. Analogous 
 plots are available for each tool. In Figure \ref{fig:toolEx} (left), the green region is the user-defined ratings for the six aspects and the overall score of the tool. The red region is the predicted ratings for the overall score and aspects our algorithm  
 \vspace{-10pt}
\vspace{-8mm}
 \noindent filled in for missing user responses. Importantly, our predicted values fall well within in the middle quartiles, indicating that our predictions are not skewing the overall tool ratings.
 In Figure \ref{fig:toolEx} (right), we compare the results of all of the tools to the responses for the individual tool. The green/red dots are the user-defined/predicted scores.
 
\vspace{-2mm}
\begin{wraptable}[20]{l}{0.4\textwidth}
\centering
\setlength \extrarowheight{4pt}
\begin{tabular}{l c}
\toprule
{\color[HTML]{000000} \textbf{Model}}  & {\color[HTML]{000000} \textbf{MSE}} \\ \hline
SGD & 0.2733 \\ 
Bayesian Ridge    & 0.2735\\ 
Kernel Ridge      & 0.2736 \\ 
Linear Regression            & 0.2762 \\ 
Support Vector    & 0.3567 \\ 
Random Forest     & 0.3786  \\ 
Gradient Boost    & 0.3887 \\ 
CatBoost          & 0.4375  \\ 
AK Modeling & 0.7379  \\ 
Elastic Net       & 0.7623 \\ 
\bottomrule
\end{tabular}
\captionsetup{font = scriptsize}
\caption{Ten machine learning algorithms compared based on Cross Validation Mean Squared Error.}
\label{tab:ml_scores}
\end{wraptable}
After having a complete user profile, we used 10 machine learning regression algorithms to predict an overall tool score (Table \ref{tab:ml_scores}). Because Stochastic Gradient Descent Regression had the highest accuracy, we used this algorithm to complete the missing values in the overall scores.

\begin{wrapfigure}[10]{r}{0.5\textwidth}
    \vspace{-30pt}
    \centering
    \includegraphics[scale = 0.27]{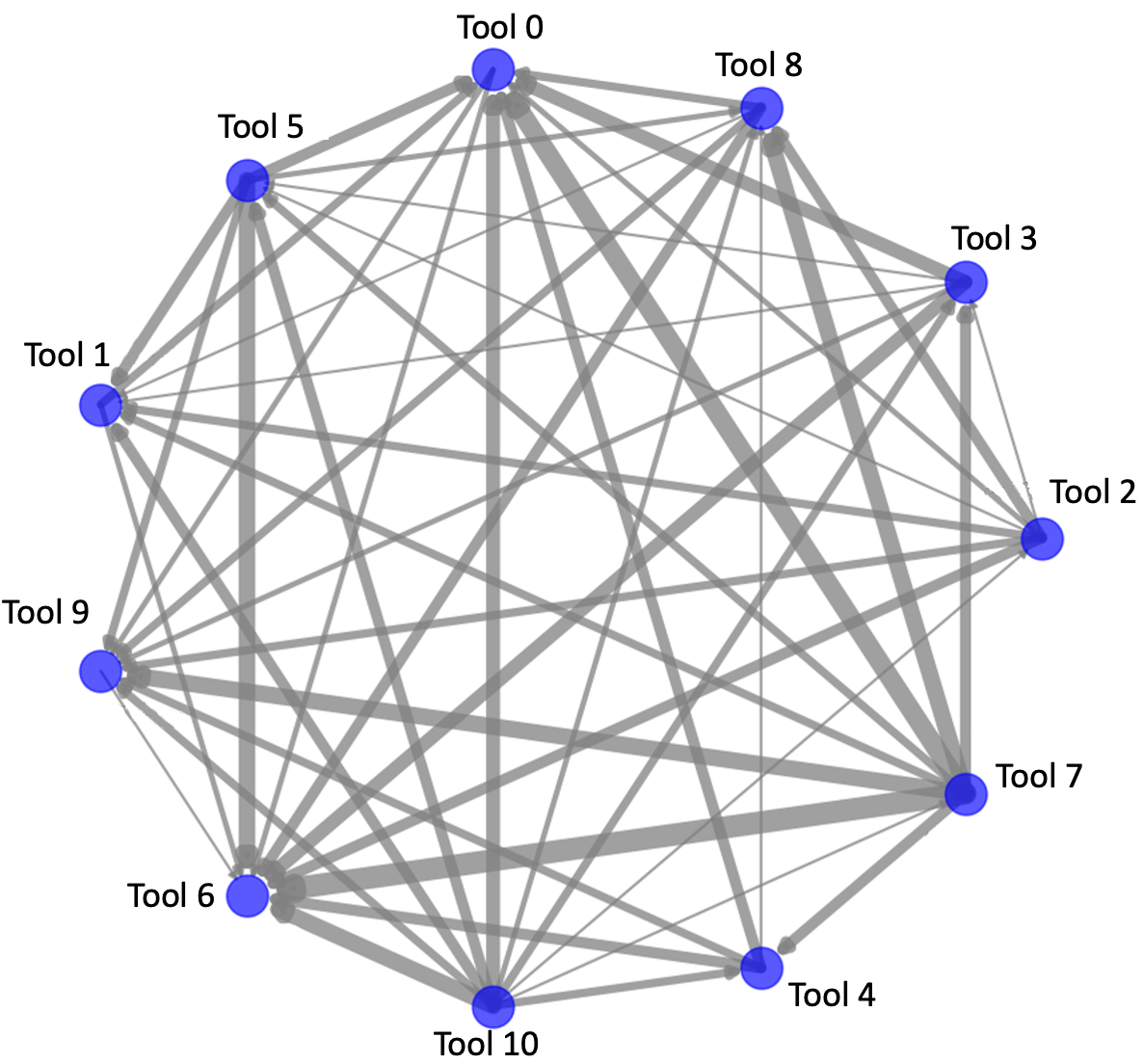}
    \captionsetup{font = scriptsize}
    \vspace{-10pt}
    \caption{PageRank graphical network from populated data based on ML algorithms.}
    \label{fig:MLresults}
\end{wrapfigure}

As previously mentioned, some users \textit{rank} tools differently then how they have \textit{rated} them, and we implement PageRank to account for these discrepancies (see fig \ref{fig:MLresults}). As we did on the raw data with missing values, we create the same directed graph for the populated data to identify the most preferred tool using only rankings. Following the pipeline we now have a complete picture of user overall scores of a tool and how that translates to user tool rankings. When calculating the overall rankings with our predicted data, we assume users rank tools based on the overall score they give a tool. If a user gives 2 tools the same overall score we then deflect to the initial tool ranking. As with the raw data, we find Tool 6 to be the 

most preferred tool and Tool 7 to be the least preferred tool using the populated data. 

\subsubsection{Statistical Analysis of Demographic Impact}\label{correlation}
\begin{figure}
    \vspace{-15pt}
    \centering
    \includegraphics[scale = 0.4]{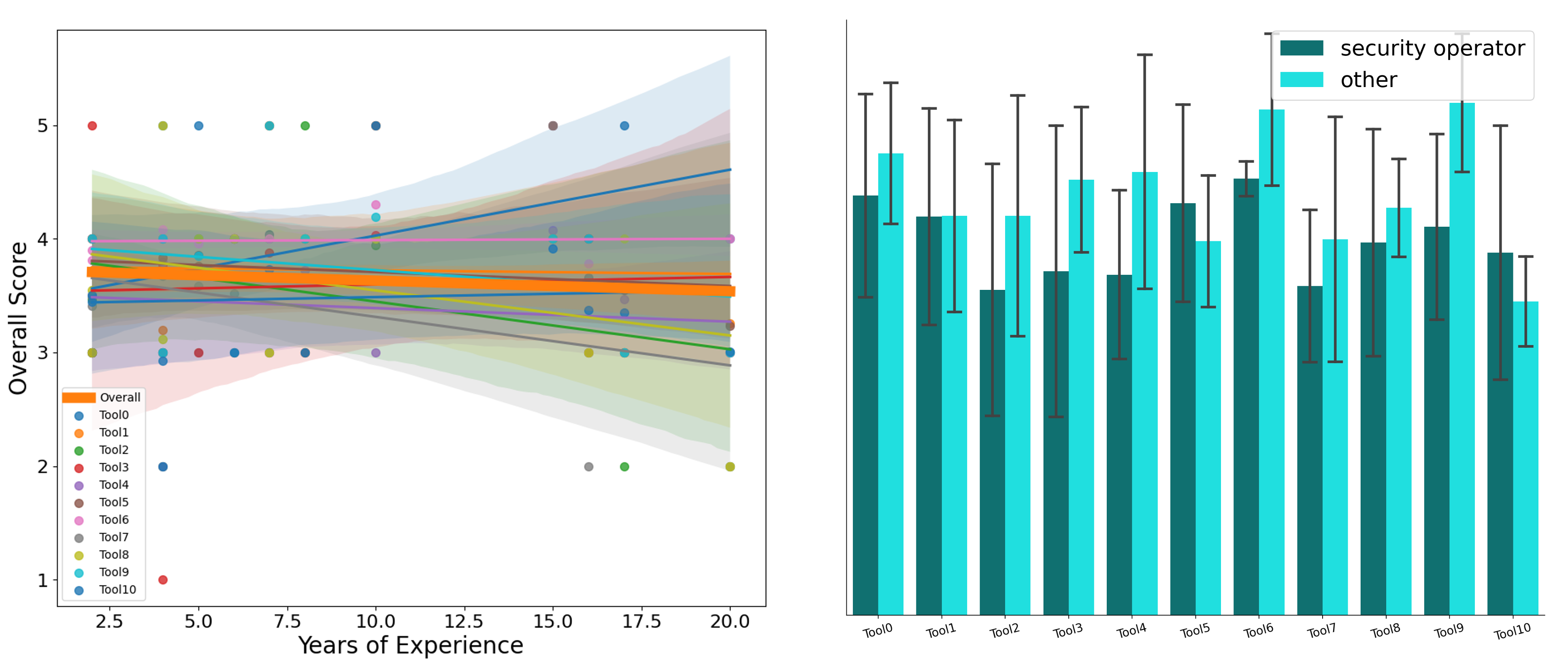}
    \captionsetup{font = scriptsize}
    \caption{There is no relationship between operator years of experience or occupation and how a tool was rated. This confirms that there is no decision bias based on experience or occupation, and even operators with similar backgrounds have different visions for SOCs.}
    \label{fig:exp}
\end{figure}

In this section we present the results related to user demographics and their correlation to overall tool rating. We selected these demographics--- years of experience, occupation, and tool familiarity ---, along with one factor pertaining to video quality, because these demographics are most likely to influence tool ratings and rankings. First, we note in Figure \ref{fig:exp} that there is no relationship between a tool's rating and the years of experience a user has. Similarly, we find that a user's occupation has no impact on tool rating, with no discernible preference for one tool over another by users of specific occupation. As such, we need not consider the effects of these factors moving forward.

\begin{wrapfigure}[15]{t!}{0.65\textwidth}
    \vspace{-18mm}
    \centering
    \includegraphics[scale = 0.27]{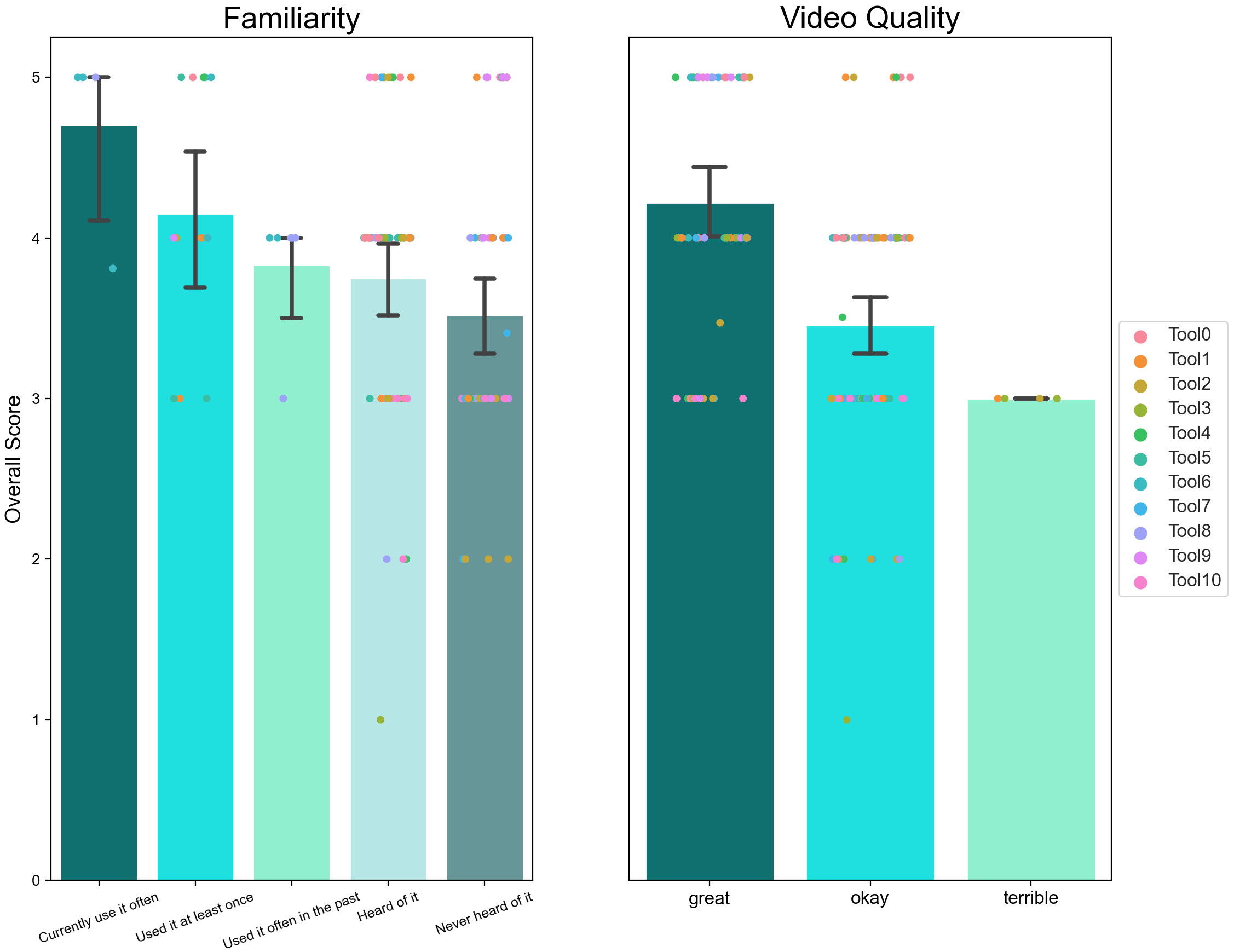}
    \captionsetup{font = scriptsize}
    \caption{In the left panel we see there is a relationship between overall score and user familiarity. In the right panel we see a relationship between video quality and overall score.}
    \label{fig:sig}
\end{wrapfigure}

However, we did identify two factors that did correlate to how a tool was rated (Figure \ref{fig:sig}). The first factor we found that influences tool rating is the familiarity a user has with a tool. We found that tools were generally rated higher by users when the user had used the tool before. The second of these was perceived quality of the submitted video. In this case, tools with higher quality videos were subsequently rated higher. It is not necessarily true that a low quality video is related to a tool's performance, and caution must be taken to ensure that the best tools are selected to move to Phase 2. Similarly, we need to account for the fact that if an operator is familiar with a tool then they likely will rate it higher.

\vspace{-2mm}
\subsection{Overall Results}
\begin{wrapfigure}[17]{l}{0.58\textwidth}
    \vspace{-6mm}
    \centering
    \includegraphics[scale = 0.5]{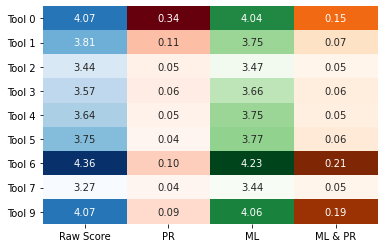}
    \captionsetup{font = scriptsize}
    \vspace{-10pt}
    \caption{Heat map summary of all four methods used to report or derive overall ratings of the tools. Raw score: average of user-defined ratings; PR: average of overall ratings derived from PageRank algorithm on the raw data; ML: average of overall ratings derived from machine learning predictions on populated data; ML + PR: average of overall ratings derived from machine learning and PageRank algorithm on populated data.}
    \label{fig: Heatmap}
\end{wrapfigure}

\vspace{-2mm}
We have two sets of data, the raw data with missing values and the populated data, and we have two ways of evaluating tool preferences, ratings and rankings. As previously mentioned, there were some discrepancies in how a user rated the tool and how a user ranked the same tool. Here, we present the results of four analyses we implemented to determine tool preferences (Figure \ref{fig: Heatmap} and Table \ref{tab:Overall Scores}). Note that the first two columns are from the raw data, and the last two columns are from the populated data. 

We note that in every analysis, Tool 7 was the bottom performer, and for

\noindent three out of four analyses, we note that Tool 2 is the the second to last performer. Similarly, Tool 6 is the top performer in three out of four cases, and only dropping to a lowest preference of third place in the PageRank analysis on raw data. It is worth noting, too, that for three of the four analyses, we have the same tools in the top three places: Tool 0, Tool 6, and Tool 9. The middle placements shuffle around considerably, but the top performers remain consistent. 
\vspace{-5mm}
\section{Discussion}
\vspace{-4mm}
In this work, we (1) provide a summary of features operators value in a SOAR tool, and (2) develop a method for down-selection when survey data is limited. Our multi-faceted approach for downselection is specifically applicable here, for progressing a subset of tools for secondary testing. This approach was centered around analysis of survey results collected from SOC operators after they viewed demonstration and overview videos on the SOAR tools. The goal of the survey was to limit thorough testing to only tools that contained many of the features that SOC operators require. 

Prior to data collection, we ran simulations to obtain an estimate for how much data and how many participants we needed to have statistically valid comparisons. Because not every operator rated every tool, machine learning was used to fill in the missing data and generate a fully populated dataset. Care was taken to address possible demographic impact, such as operator occupation and years of experience. 

We note that the two most important features operators are looking for in a SOAR tool are (1) its ability to automate common tasks and (2) functionality of playbooks. More than half of the participants voted that playbooks were the most important aspect of a SOAR tool, followed by task automation, ticketing, and ranking of alerts in a clear second, third, and fourth order, respectively. This survey, even with its limitations, provides a starting point for SOAR tool vendors to focus their efforts on aspects that are most beneficial to SOC function.

Another survey aspect that affected our analysis of operator preferences were the discrepancies between how an operator rated tools vs ranked tools. For example, in many instances operators would rate Tool 0 at a 5/5 and Tool 1 at a 3/5, but then would rank Tool 1 better than Tool 0. Due to the numerous discrepancies between ranking and rating we implemented the PageRank analysis algorithm to develop a scoring metric for the ratings. This algorithm allowed us to predict which node is the most preferred even with the user discrepancies.

Because participants viewed videos created by SOAR tool vendors rather than using the tools directly, it is possible that their opinions were influenced by either video quality or prior knowledge of a particular tool.
We asked participants to provide their opinion of video quality and their prior experience with each tool in our survey, and we discuss correlations between these factors and our results in Section~\ref{correlation}. 
It is also possible that their opinion of the tool would change if they had the opportunity to use it in an operational context, which is the next phase of our research. However, the largest limitation to our study was the number of participants. We recognize that in an ideal world we would have had more participants, however given the information we found about how many participants would provide what confidence, our sponsor determined they were satisfied with nineteen for this phase of down-selecting. 

Based on this analysis, we are equipped to identify a subset of these 11 tools that will be thoroughly tested. Operators will use these tools in several realistic scenarios and will be asked to complete a survey after regarding their experience. Furthermore specific performance metrics will be collected to measure the improvement to efficiency and effectiveness by usage of SOAR tools in a SOC. While this framework was developed in the context of SOC survey responses and downselecting a large sample, the impacts of this study are far-reaching. In general the framework we present provides designers of user studies with the ability to quantify the statistical power of their analyses based on their sample size, and furthermore provides a reliable method of populating missing user data. If there are multiple methods of evaluation, such as scores and ranking, we demonstrate a novel method of reconciling differences between the two for a more clear interpretation and meaningful results.

\appendix
\section{Appendix}
\subsection*{Acknowledgements}
Special thanks to Jeff Meredith for assisting with the website. 
The research is based upon work supported by the Department of Defense (DOD), Naval Information Warfare Systems Command (NAVWAR), via the Department of Energy (DOE) under contract  DE-AC05-00OR22725. The views and conclusions contained herein are those of the authors and should not be interpreted as representing the official policies or endorsements, either expressed or implied, of the DOD, NAVWAR, or the U.S. Government. The U.S. Government is authorized to reproduce and distribute reprints for Governmental purposes notwithstanding any copyright annotation thereon. 
\\
\begin{table}
\centering
\setlength\extrarowheight{3pt}
\begin{tabular}{m{2.95cm} m{2.75cm} m{2.75cm} m{2.75cm}}
\toprule
\textbf{Raw Score} $[1-5]$ & \textbf{PR} $[0-1]$ & \textbf{ML} $[1-5]$ & \textbf{ML + PR} $[0-1]$\\
\hline
Tool 6 (4.363)  & Tool 0 (.341) & Tool 6 (4.23)  & Tool 6 (.207) \\ 
Tool 9 (4.071) & Tool 1 (.114)    & Tool 9 (4.06) & Tool 9 (.192) \\ 
Tool 0 (4.071)  & Tool 6 (.097)  & Tool 0 (4.04) & Tool 0 (.146)  \\ 
Tool 1 (3.813) & Tool 9 (.085) & Tool 5 (3.77)   & Tool 1 (.070)  \\ 
Tool 5 (3.750) & Tool 3 (.057)  & Tool 4 (3.75)  & Tool 5 (.061)   \\ 
Tool 4 (3.636)  & Tool 2 (.052) & Tool 1 (3.75)  & Tool 3 (.055)  \\ 
Tool 3 (3.571)  & Tool 4 (.048)  & Tool 3 (3.66) & Tool 4 (.054)  \\ 
Tool 2 (3.438)  & Tool 5 (.043)  & Tool 2 (3.47) & Tool 2 (.046)  \\
Tool 7 (3.273)  & Tool 7 (.042) & Tool 7 (3.44) & Tool 7 (.046)  \\ 
\bottomrule
\end{tabular}
\captionsetup{font = scriptsize}
\caption{Four methods used to report or derive overall ratings of the tools. Raw score: average of user-defined ratings; PR: average of overall ratings derived from PageRank algorithm on the raw data; ML: average of overall ratings derived from machine learning predictions on populated data; ML + PR: average of overall ratings derived from machine learning and PageRank algorithm on populated data.}
\label{tab:Overall Scores}
\end{table}


\begin{center}
\begin{table}
\setlength\extrarowheight{4pt}
\begin{tabular}{  m{1.85cm}  m{2.5cm} m{8cm}  } 
 \toprule
  \textbf{Question} \# & \textbf{Question type} & \textbf{Question} \\
  \hline
  1 & Pre-Survey & How familiar are you with SOAR tools?\\ 
  2 &  Pre-Survey &  Which of these best fits your role?   \\ 
  3 & Pre-Survey & How many years have you been in that role? \\ 
  4 & Pre-Survey & Please rank the following capabilities in order of importance, with 1 being the most important and 7 being the least important in your SOC. \\ 
  \midrule
  1 & Familiarity & How familiar are you with this tool? \\
  2 & Quality & What do you think of the quality of these videos? \\ 
  3 & Overall Score & What is you overall impression of this tool? \\ 
  4 & Ranking & Does the tool present and prioritize data in a way that is beneficial? \\ 
  5 & Ingestion & Do you think this tool could effectively ingest the data in your SOC? \\ 
  6 & Playbooks & Does the tool provide steps (playbook, workflow) that guide tier 1 or junior analysts through common tasks? \\ 
  7 & Ticketing & Does the tool automate tasks in a way that would increase efficiency? \\ 
  8 & Collaboration & Does the tool enable multiple analysts to effectively collaborate (simultaneously)? \\ 
  9 & Automation & Does the tool enable a hand off of investigations (for example, between two shifts or across SOCs)? \\
  10 & Free response & Is there anything else about this tool that you would like to share? \\
  N/A & Overall Ranking & Please rank the tools that you reviewed by order of preference, where 1 indicates the tool that you would most like to see used in your SOC. You can drag and drop the tool names to reorder them. \\
  \bottomrule
  \end{tabular}
  \captionsetup{font = scriptsize}
  \caption{\label{tab:Survey}Survey questionnaire given to the SOC operators. The survey was delivered electronically and included 4 pre-survey questions, 10 questions about the specific tools (including their aspects), and 1 question about the overall ranking. All ratings questions were scored 1-5, with 1 being the worst and 5 being the best. On the two ranking questions, the operators ranked their most preferred aspect/tool as 1 and their least preferred as the highest value.}
  \end{table}
\end{center}

\bibliographystyle{IEEEtran}
\bibliography{bib/refs,bib/toolrefs}
\end{document}